\documentclass[12pt]{article} 
\usepackage[xdvi]{graphicx} 
\usepackage{amssymb}
\usepackage{amsmath} 
 
\begin{document}   
\newcommand{\todo}[1]{{\em \small {#1}}\marginpar{$\Longleftarrow$}}   
\newcommand{\labell}[1]{\label{#1}\qquad_{#1}} 

\rightline{DCPT-04/05}   
\rightline{gr-qc/0402044}   
\vskip 1cm


\begin{center} 
{\Large \bf On Gauss-Bonnet black hole entropy}
\end{center} 
\vskip 1cm   
  
\renewcommand{\thefootnote}{\fnsymbol{footnote}} \centerline{\bf
Tim Clunan\footnote{T.P.Clunan@durham.ac.uk}, Simon
F. Ross\footnote{S.F.Ross@durham.ac.uk} and Douglas J.
Smith\footnote{Douglas.Smith@durham.ac.uk}}
\vskip .5cm   
\centerline{ \it Centre for Particle Theory, Department of  
Mathematical Sciences}   
\centerline{\it University of Durham, South Road, Durham DH1 3LE, U.K.}   
 
\setcounter{footnote}{0}   
\renewcommand{\thefootnote}{\arabic{footnote}}


\begin{abstract}   
We investigate the entropy of black holes in Gauss-Bonnet and Lovelock
gravity using the Noether charge approach, in which the entropy is
given as the integral of a suitable $(n-2)$ form charge over
the event horizon. We compare the results to those obtained in other
approaches. We also comment on the appearance of negative entropies in
some cases, and show that there is an additive ambiguity in the
definition of the entropy which can be appropriately chosen to avoid
this problem.  
\end{abstract}    
  
\section{Introduction}

There has recently been considerable interest in the study of
higher-curvature corrections to the Einstein-Hilbert action,
particularly in the context of brane
worlds~\cite{bg1,bg2,Germani1,bg3,bg4,bg5,bg6,bg7a,bg7,bg8,Germani2,Germani3,bg9,bg10,ruth}. These
terms are important because their presence can lead to qualitative
changes in the physics as seen from the point of view of observers on
the brane (see, e.g.,~\cite{bg10,ruth}). In addition, such terms appear
in the low-energy effective theories obtained from string
theory~\cite{Zwiebach,Nepomechie}; it is thus very natural to include
them in the context of models motivated by string theory.

Typically, only certain special corrections are considered. At
quadratic order, the combination which is considered is the
Gauss-Bonnet term,
\begin{equation}
{\cal L} = R_{\alpha\beta\gamma\delta} R^{\alpha\beta\gamma\delta} - 4
R_{\alpha\beta} R^{\alpha\beta} + R^2.
\end{equation}
In four dimensions this is a topological invariant; in higher
dimensions, it is the most general quadratic correction which
preserves the property that the equations of motion involve only
second derivatives of the metric. This combination is thus
particularly tractable, and explicit solutions to the resulting
equations of motion have been found. It has also been shown that this
gives a useful Lagrangian description of the leading correction to the
low-energy effective action in heterotic string
theory~\cite{Zwiebach,Nepomechie}. The special nature of the
Gauss-Bonnet term is more apparent when it is written in a
differential forms notation; it is simply
\begin{equation}
\mathbf{L} = \epsilon_{a_1 \ldots a_n} \mathbf{R}^{a_1 a_2} \mathbf{R}^{a_{3}
  a_4}  \mathbf{e}^{a_5}  \cdots  \mathbf{e}^{a_n}, 
\end{equation}
where $n$ is the spacetime dimension, $\mathbf{e}^{a}$ is a vielbein
one-form, $\mathbf{e}^{a} = e^a_{\ \alpha} dx^\alpha$, and $\mathbf{R}^{a_1
  a_2}$ is the curvature two-form, which is related to the Riemann
tensor by $R^{ab}_{\ \ \gamma\delta} = R^{\alpha\beta}_{\ \
  \gamma\delta} 
e^{a}_{\ \alpha} e^{b}_{\ \beta}$.

At higher orders, the natural generalisation is to consider the
dimensionally continued Euler densities  
\begin{equation}
\mathbf{L}^{(p)} = \epsilon_{a_1 \ldots a_n} \mathbf{R}^{a_1 a_2}  \cdots
\mathbf{R}^{a_{2p-1} a_{2p}}  \mathbf{e}^{a_{2p+1}}  \cdots   \mathbf{e}^{a_n}.
\end{equation}
The first $\mathbf{L}^{(1)}$ is just the Einstein-Hilbert action.  The
Lagrangian density constructed from Einstein gravity plus these higher-order
corrections defines the Lovelock
theory~\cite{Lovelock:1971}:
\begin{equation} \label{lll}
\mathbf{L}=\sum_{p=0}^{\lfloor n/2 \rfloor} \alpha_p \mathbf{L}^{(p)}.
\end{equation}
This is the most general Lagrangian invariant under local Lorentz
transformations, constructed from the vielbein, the spin connection,
and their exterior derivatives, without using the Hodge dual; such
that the field equations for the metric are second order
\cite{Zumino:1986,Regge:1986,Teitelboim:1987}. We will only consider
in detail the choice of $\alpha_p$ discussed
in~\cite{Banados:1994,bhscan}, which gives a unique cosmological
constant (i.e., a unique constant-curvature vacuum solution).

Since the equations of motion are of second order, it is possible to
find explicit solutions, and black hole solutions of these theories
have been found by several
authors~\cite{Boulware:1985,Wheeler:1986,Myers:1988,Wiltshire:1988,Banados:1994,Cai:1998,bhscan,Aros:2000,Cai:2001}.
In the context of brane-world models, these black hole solutions play
an important role, as non-trivial cosmological evolutions are obtained
by considering the motion of a brane in a bulk black hole solution.
The higher-order corrections in the action lead to modifications in
the formula for the entropy of these black hole solutions; the entropy
is no longer simply given by the area of the black hole's event
horizon.  The entropy of the black hole solutions was calculated
in~\cite{Myers:1988,bhscan,Cai:2001}. In~\cite{Myers:1988}, the
Euclidean approach was used, while in~\cite{bhscan,Cai:2001} the
entropy was calculated from the other thermodynamic quantities using
the first law of thermodynamics. The thermodynamics of these solutions
was further studied in~\cite{Neu1,Neu2,Neu3}, with the same results
for the entropy. 

Our aim in the present paper is to investigate the entropy of these
black hole solutions using the general approach to black hole entropy
for arbitrary Lagrangian theories introduced by Iyer and
Wald~\cite{Wald,Iyer:1994,Iyer:1995}\footnote{In fact, the special
case of Lovelock gravity of interest here was investigated earlier by
similar techniques in~\cite{JacMy}.}. This is referred to as the
Noether charge approach, because the entropy is always expressed in
terms of the integral of an $(n-2)$-form charge $\mathbf{Q}$ over the
event horizon.  This approach thus extends the relation between
entropy and geometry from general relativity to arbitrary metric
theories of gravity. The expressions for the entropy we obtain from
the Noether charge approach agree with those obtained
in~\cite{Myers:1988,bhscan,Cai:2001}.  Since the Noether charge
entropy satisfies the first law of thermodynamics by construction,
this is unsurprising. In fact, one can think of this Noether charge
approach as a more formal version of the integration of the first law
used in~\cite{bhscan,Cai:2001}. However, by using this approach, we
will gain some insight into the geometric significance of these
expressions for the black hole entropy. Similar results were obtained
by the conical singularity method in~\cite{Furs}.

We will also address the appearance of negative entropies in these
calculations. In~\cite{Od1,Od2,Cai:2003}, it was observed that the
entropy assigned to some solutions could be negative.
In~\cite{Cai:2003}, it was suggested that some of these solutions
which are assigned negative entropy should be discarded as unphysical.
We will argue instead that this occurrence of negative entropies
reflects a deficiency in the methods used to calculate the entropy. As
noted also in~\cite{Od2}, there is an additive ambiguity in the
definition of entropy in the approach used
in~\cite{bhscan,Cai:2001,Cai:2003}; one can add an arbitrary constant
to the entropy of all the solutions in some family of solutions
without affecting the first law. By changing the choice of this
constant, one can clearly arrange to have all solutions have positive
entropy. That is, we have to decide which classical solution we assign
zero entropy; if we make this choice appropriately, all solutions have
positive entropy.

We had initially hoped that using the Noether charge approach would
resolve this ambiguity. However, since the results we obtain using the
Noether charge definition of the entropy
from~\cite{Wald,Iyer:1994} agree with the previous calculation,
this cannot be true. In fact, we will see that there is a
corresponding ambiguity in this approach: we can redefine $\mathbf{Q}$
by adding to it a closed but not exact $(n-2)$-form. The existence of
such a form is a consequence of the non-trivial topology of the black
hole solutions.  Thus, the negative entropies are again removed by an
appropriate choice of zero entropy.

In the next section, we will review the black hole solutions we will
consider, and describe the thermodynamic results obtained
in~\cite{bhscan,Cai:2001}. In section~\ref{noether}, we will
briefly review the Noether charge approach to black hole entropy, and
apply it to these solutions. In section~\ref{neg}, we point out that
some solutions have been assigned negative entropy, and explain how we
can correct this problem by changing our choice of zero entropy
solution in both approaches.

\section{Black hole solutions}
\label{bh}

Black hole solutions of a theory with higher-curvature corrections to
the Lagrangian were first written down in~\cite{Boulware:1985}, where
the first three terms in the Lagrangian (\ref{lll}) were kept: that
is, the usual action for Einstein gravity with a cosmological constant
was modified by the addition of the Gauss-Bonnet
term, 
\begin{equation} \label{gblag}
I = {1 \over 16 \pi G} \int d^n x \sqrt{-g} \left( R - 2 \Lambda +
\alpha ( R_{\alpha\beta\gamma\delta} R^{\alpha\beta\gamma\delta} - 4
R_{\alpha\beta} R^{\alpha\beta} + R^2) \right).
\end{equation}
We will sometimes write the cosmological constant as $\Lambda = -
(n-1)(n-2)/2l^2$, since the interest from the brane world point of
view is mainly in the case where the cosmological constant is
negative. 
 
Attention was restricted to static solutions, so the ansatz for
the metric is
\begin{equation} \label{bhmet}
ds^2 = -F^2(r)dt^2+\frac{1}{G^2(r)}dr^2+r^2 h_{\mu \nu} dx^{\mu}
dx^{\nu},
\end{equation}
where $h_{\mu\nu}$ is a positive definite metric independent of $r,t$,
and $\mu,\nu$ run over $n-2$ values. In~\cite{Boulware:1985},
spherical symmetry was considered, so $h_{\mu\nu}$ was taken to be the
metric on an $S^{n-2}$. In~\cite{Cai:2001}, this was extended to
consider $h_{\mu\nu}$ a metric of constant curvature $(n-2)(n-3)k$
where $k=0,\pm 1$. We denote the volume of this space by $\Sigma_k$.
The physically relevant solution of the equations of motion has
\begin{equation} \label{gbF}
F^{2}(r)=G^2(r)= k+\frac{r^2}{2 \tilde{\alpha}} \left(1- \sqrt{1+\frac{64 \pi
G \tilde{\alpha} M}{(n-2) \Sigma_k r^{n-1}}+{8 \tilde \alpha \Lambda \over 
 (n-1)(n-2)}} \right),
\end{equation}
where $\tilde\alpha = (n-3)(n-4) \alpha$.  Since horizons occur where
$F^2(r)=0$, we see that black hole solutions with $k=0$ or $k=-1$ are
only possible when the cosmological constant is negative. For negative
or zero cosmological constant, the largest root of $F^2(r)=0$ defines
the black hole event horizon $r=r_+$. For positive cosmological
constant, we take $k=+1$, and the largest root is the cosmological
horizon $r=r_c$, and the next root is the event horizon $r=r_+$. 

In~\cite{Cai:2001}, the mass and temperature of these black hole
solutions was obtained in the usual way. The parameter $M$ in the
above solution is the mass, which can be expressed in terms of the
horizon radius $r_+$ by
\begin{equation}
M = {(n-2) \Sigma_k r_+^{n-3} \over 16 \pi G} \left( k+ {\tilde \alpha
\over r_+^2} + {r_+^2 \over l^2} \right). 
\end{equation}
We will assume that $M \geq 0$. In~\cite{Neu1,Neu2}, it was argued
that the energy for the $k=-1$ case differs from this mass by a
background subtraction; this does not affect the analysis of the
entropy. The temperature is obtained by determining the natural
periodicity in the Euclidean time direction, giving
\begin{equation}
T = {(n-1) r_+^4 + (n-3) k l^2 r_+^2 + (n-5) \tilde \alpha k^2 l^2
\over 4\pi l^2 r_+ (r_+^2 + 2 \tilde \alpha k)}. 
\end{equation}
These expressions were then used to obtain an expression for the
entropy by integrating up the first law. That is, we assume that the
family of black hole solutions parametrised by $M$, or alternatively
$r_+$, satisfies the first law of thermodynamics $dM = T dS$, and we
obtain an expression for the entropy by integrating this relation over
the family of solutions,
\begin{equation} \label{therment}
S = \int T^{-1} dM = \int_0^{r_+} T^{-1} \left( {\partial M \over
\partial r_+} \right) dr_+. 
\end{equation}
In the second step, the choice of lower limit of integration
represents the additive ambiguity in the definition of the entropy
referred to in the introduction. This specification of lower limit
represents the reasonable assumption that, as in the usual formula for
the entropy in Einstein gravity, we should take the entropy to go to
zero when the area of the horizon vanishes. This gives a formula
\begin{equation} \label{gbent}
S = \frac{r_{+}^{n-2} \Sigma_k}{4 G} \left(1+\frac{2 \tilde{\alpha} k
(n-2)}{(n-4) r_{+}^{2}}\right) 
\end{equation}
for the entropy of these black hole solutions. 

We wish to point out two features of this black hole entropy formula.
First, even though the entropy was obtained purely by thermodynamic
arguments, it has evident relations to the geometry of the event
horizon: in the case $k=0$, where the horizon is flat, the
Gauss-Bonnet term has no effect on the expression for the entropy,
which is simply the area of the event horizon. Furthermore, the
correction term is controlled in general by the curvature of the event
horizon. In the present approach, it is difficult to see why this
should be the case.  In the next section, we will use the Noether
charge approach to obtain a more geometrical understanding of this
observation. Secondly, we note that the entropy is not necessarily
positive. As observed in~\cite{Od1,Od2,Cai:2003}, if $\alpha k <0$,
the second term in (\ref{gbent}) is negative, and this term can make
the whole expression negative for sufficiently small black holes. We
will explore this observation in more detail in section~\ref{neg}.

The extension to consider black hole solutions when further
corrections are turned on was explored
in~\cite{bhscan,Cai:1998,Aros:2000}, where the full Lovelock
Lagrangian (\ref{lll}) was considered for particular choices of the
coefficients $\alpha_p$, 
\begin{equation} \label{llalpha}
\alpha_p = {l^{2(p-q)} \over 2(n-2p)(n-2)! \Omega_{n-2} G_q}
\left(\begin{array}{c}q \\ p \end{array}\right) 
\end{equation}
for $p \leq q$ and zero for $p>q$, where $q$ is an integer less than
or equal to $(n-1)/2$.\footnote{The entropy for the more general
  Lovelock Lagrangian was studied in~\cite{Wu,Cai:ll}. For simplicity,
  we will restrict to the above choices.} The static spherically symmetric black hole
solutions of the resulting theories were found in~\cite{bhscan}; these
were extended to non-trivial horizon topology
in~\cite{Cai:1998,Aros:2000}. They are given by the same ansatz
(\ref{bhmet}), with $h_{\mu\nu}$ a metric of constant curvature
$k = 0, \pm 1$ and
\begin{equation} \label{llF}
  F^2(r) =G^{2}(r)=k +\frac{r^2}{l^2}-\left(\frac{2 G_{q} M+
  \delta_{n-2q,1}}{r^{n-2q-1}}\right)^{1/q}. 
\end{equation}
The entropy of these black hole solutions was found by a similar
argument to the one employed in the Gauss-Bonnet example; the result
is
\begin{equation} \label{llent}
S = \frac{2\pi q}{G_q} \int_{0}^{r_{+}} r^{n-2q-1}
(k+\frac{r^2}{l^2})^{q-1} dr. 
\end{equation}
Here, we once again see that for $k=0$, the correction to the
usual entropy formula drops out; more generally, the correction is
determined by the curvature of the horizon. In the next section, we
will also reproduce this result from the Noether charge point of view.

\section{Black hole entropy as a Noether charge}
\label{noether}

In general relativity, the black hole entropy is directly connected to
the geometry: it is simply the area of the event horizon in
appropriate units. This relation forms part of a very strong relation
between thermodynamics and geometry, particularly in the Euclidean
approach to black hole thermodynamics, where the black hole entropy
can be seen to arise from the non-trivial topology of the Euclidean
black hole solutions. As remarked above, the black hole entropy in
higher-curvature theories of gravity takes a more complicated form. It
would clearly be very interesting to understand the geometrical origin
of the expressions (\ref{gbent},\ref{llent}).

In~\cite{Wald,Iyer:1994}, a geometrical expression for the black
hole entropy in an arbitrary diffeomorphism-invariant theory of
gravity was obtained, and shown to satisfy the first law in
general. This approach is based on a Noether current constructed from
a diffeomorphism variation of the Lagrangian.

We consider an arbitrary Lagrangian density $\mathbf{L}$ depending on
a collection of dynamical fields $\phi = (g_{\mu\nu},\psi)$, where
$\psi$ represents matter degrees of freedom. The variation of the
Lagrangian density under a general variation $\delta \phi$ of the
dynamical fields was shown to be given by an equation of motion term
plus a surface term, 
\begin{equation}
\delta \mathbf{L} = \mathbf{E} \delta \phi + d
\mathbf{\Theta}(\phi,\delta \phi).
\end{equation}
A Noether current was then defined by 
\begin{equation}
\mathbf{J} = \mathbf{\Theta}(\phi, \mathcal{L}_\xi \phi) - \xi \cdot
\mathbf{L},
\end{equation}
for $\xi^\mu$ some smooth vector field. That is, we take the surface
term in the variation when the infinitesimal variation arises from a
diffeomorphism, $\delta \phi = \mathcal{L}_\xi \phi$, and subtract the
vector field contracted with $\mathbf{L}$. This current was shown to
be a closed form on-shell, so we can define an $(n-2)$-form charge
$\mathbf{Q}$ by $\mathbf{J} = d\mathbf{Q}$. 

This Noether charge $\mathbf{Q}$ then plays the central role in
defining the entropy. It can always be expressed as
\cite{Iyer:1995}\footnote{Here $
\nabla_{[a} \xi_{b]} = e^{\ \alpha}_a e^{\ \beta}_b \nabla_{[\alpha}
\xi_{\beta]}$.}
\begin{equation}
\mathbf{Q} = \mathbf{W}_a(\phi) \xi^a + \mathbf{X}^{ab}(\phi)
\nabla_{[a} \xi_{b]} + \mathbf{Y}(\phi,\mathcal{L}_\xi \phi) + d
\mathbf{Z} (\phi, \xi),
\end{equation}
where
\begin{equation}
\mathbf{X}_{ab} = - {\delta \mathbf{L} \over \delta \mathbf{R}^{ab}}, 
\end{equation}
that is, $\mathbf{X}_{ab}$ is equal to the variation of the Lagrangian with
respect to the curvature, if we were to treat the curvature as an
independent field. 
This expression for the Noether charge is clearly not unique;
the form of $\mathbf{Z}$ is completely arbitrary, and there are other
ambiguities arising from ambiguities in the definitions of
$\mathbf{L}$ and $\mathbf{\Theta}$. 

It was then proposed in~\cite{Iyer:1994,Iyer:1995} that we define the
entropy of an asymptotically flat black hole solution with a bifurcate
Killing horizon by
\begin{equation} \label{nent}
S = 2\pi \int_{\Sigma} \mathbf{Q}[t] = 2\pi \int_{\Sigma}
\mathbf{X}^{cd} \epsilon_{cd},
\end{equation}
where the integral is over the bifurcation $(n-2)$-surface in the
black hole solution, and $t$ is the horizon Killing field, normalised
to have unit surface gravity. In the second relation, we have used the
fact that $t$ is a Killing vector which vanishes on $\Sigma$, and
$\epsilon_{cd}$ is the binormal to the surface $\Sigma$. The main
result of~\cite{Iyer:1994,Iyer:1995} is that this entropy
automatically satisfies the first law of thermodynamics,
\begin{equation}
{\kappa \over 2\pi} \delta S = \delta \mathcal{E} - \Omega_H^{(\mu)}
\delta \mathcal{J}_{(\mu)},
\end{equation}
where $\kappa$ is the surface gravity and $\mathcal{E}$ and
$\mathcal{J}_{(\mu)}$ are energy and angular momenta defined in terms
of surface integrals at infinity. This approach thus gives a
satisfactory definition of the black hole entropy in terms of an
integral over the black hole's event horizon. An interesting attempt
to relate this general entropy formula to a microscopic description
appeared in~\cite{pal1,pal2}. 

We would now like to apply this approach to the calculation of the
entropy of the black hole solutions discussed in the previous
section. For the Einstein+Gauss-Bonnet theory, the appropriate
$\mathbf{Q}$ was worked out in~\cite{Iyer:1995}:
\begin{equation} \label{gbq}
Q_{\alpha_1 \ldots \alpha_{n-2}} = - \epsilon_{\gamma \delta \alpha_1 \ldots
  \alpha_{n-2}} \left( {1 \over 16 \pi} \nabla^\gamma\xi^\delta + 2 \alpha
  ( R \nabla^\gamma \xi^\delta + 4 \nabla^{[\beta} \xi^{\delta]}
  R^\gamma_{\ \beta} + R^{\gamma \delta \beta \kappa} \nabla_{\beta}
  \xi_\kappa )\right). 
\end{equation}
Putting this into (\ref{nent}) and evaluating the integral over the
bifurcation surface $r=r_+$ in the black hole geometry
(\ref{bhmet},\ref{gbF}) does reproduce the entropy (\ref{gbent})
obtained previously. It is pleasing to see the two approaches
agree---although hardly surprising, since (\ref{gbent}) was calculated
by assuming the entropy satisfies the first law, and (\ref{nent}) does
so. 

However, our hope was that performing the calculation in terms of the
Noether charge would offer some insight into the evident relation
between the entropy formula (\ref{gbent}) and the geometry of the
horizon. Although the Noether charge formulation allows us to express
the entropy as a surface integral over the horizon, the origin of the
$k$ dependence in (\ref{gbent}) remains somewhat obscure. To obtain
further insight into this question, we will redo the calculation in
the language of differential forms. 

To expose the geometric structure, we want to calculate the integral
(\ref{nent}) over the bifurcation surface in a general metric of the
form (\ref{bhmet}), without at first fixing the forms of $F(r)$,
$G(r)$ or $h_{\mu\nu}$. We will consider the general Lagrangian
(\ref{lll}) for the Lovelock theory. It is easy to see that
\begin{equation}
\mathbf{X}_{ab} = - \sum_{p=1}^{\lfloor n/2 \rfloor} p \alpha_p
\epsilon_{a b a_3   \ldots a_n} \mathbf{R}^{a_3 a_4}  \cdots
\mathbf{R}^{a_{2p-1} a_{2p}}  \mathbf{e}^{a_{2p+1}}  \cdots
\mathbf{e}^{a_n}.
\end{equation}
For the metric (\ref{bhmet}), a suitable choice of orthonormal basis
is 
\begin{equation}
\mathbf{e}^1=F \mathbf{dt}, \quad \mathbf{e}^2=\frac{1}{G}\mathbf{dr},
\quad \mathbf{e}^{m}=\mathbf{\tilde{e}}^{m}, 
\end{equation}
where $\mathbf{\tilde{e}}^m$, $m=3,\ldots,n$ is a suitable orthonormal
basis corresponding to the induced metric $r^2 h_{\mu\nu}$ on the $(n-2)$
surface. The spin connection computed for this vielbein has 
\begin{equation}
\pmb{\omega}^{mn} = \boldsymbol{\tilde{\omega}}^{mn}, \quad
  \boldsymbol{\omega}^{2m} = - {G \over 2r} \mathbf{\tilde{e}}^m,
\end{equation}
and hence
\begin{equation}
\mathbf{R}^{mn} = \mathbf{\tilde{R}}^{mn} - \left( {G \over 2r}
\right)^2 \mathbf{\tilde{e}}^m \mathbf{\tilde{e}}^n.
\end{equation}
We will not need the other components of the curvature
two-form. Hence
\begin{eqnarray}
\mathbf{X}_{12} &=& - \sum_{p=1}^{\lfloor n/2 \rfloor} p \alpha_p
\epsilon_{1 2 a_3   \ldots a_n} \mathbf{R}^{a_3 a_4}  \cdots
\mathbf{R}^{a_{2p-1} a_{2p}}  \mathbf{e}^{a_{2p+1}}  \cdots
\mathbf{e}^{a_n} \\
&=& - \sum_{p=1}^{\lfloor n/2 \rfloor} p \alpha_p
\epsilon_{m_1   \ldots m_{n-2}} \mathbf{R}^{m_1 m_2}  \cdots
\mathbf{R}^{m_{2p-3} m_{2p-2}}  \mathbf{e}^{m_{2p-1}}  \cdots
\mathbf{e}^{m_{n-2}}  \\
&=& - \sum_{p=1}^{\lfloor n/2 \rfloor} p \alpha_p \left(
  \mathbf{\tilde{R}}^{m_1 m_2} - \left( {G \over 2r} 
\right)^2 \mathbf{\tilde{e}}^{m_1} \mathbf{\tilde{e}}^{m_2} \right)
\cdots \left(
  \mathbf{\tilde{R}}^{m_{2p-3} m_{2p-2}} - \left( {G \over 2r} 
\right)^2 \mathbf{\tilde{e}}^{m_{2p-3}} \mathbf{\tilde{e}}^{m_{2p-2}}
\right)  \nonumber \\ && \times
\mathbf{\tilde{e}}^{m_{2p-1}}  \cdots
\mathbf{\tilde{e}}^{m_{n-2}}\\
&=& - \sum_{p=1}^{\lfloor n/2 \rfloor} p \alpha_p \sum_{j=0}^{p-1}
(-1)^{p-1-j} \left( \begin{array}{c}p-1\\j\end{array}\right)
\left(\frac{G}{2 r} \right)^{2(p-1-j)} \mathbf{\tilde{L}}^{(j)}. 
\end{eqnarray} 
We see that the term in $\mathbf{X}_{cd}$ that is relevant to
the evaluation of the entropy is a certain combination of the Euler
densities evaluated on the $(n-2)$-dimensional submanifold. This is
the key to understanding the very suggestive form of
(\ref{gbent}).\footnote{This result is also interestingly reminiscent of
  the study of codimension two braneworlds in \cite{ruth}, where it was
  found that a Gauss-Bonnet action in the bulk induced Einstein
  gravity on the brane. We do not yet fully understand the relation
  between the two results.} 

In the particular case of the Gauss-Bonnet theory (\ref{gblag}), we
can use the fact that $\alpha_p = 0$ for $p >2$ and that $G(r)$
vanishes on the event horizon in the solution (\ref{gbF}) to write
\begin{equation}
S = 2\pi \int_{\Sigma} \mathbf{X}^{cd} \epsilon_{cd} = 4\pi
\int_{\Sigma} \mathbf{X}^{12} = 4\pi
\int_{\Sigma} ( \alpha_1 \mathbf{\tilde{L}}^{(0)} + 2 \alpha_2
\mathbf{\tilde{L}}^{(1)}).
\end{equation}
That is, the entropy contains a term proportional to the area of the
event horizon (since $\mathbf{\tilde{L}}^{(0)}$ is just the horizon
volume form) and a term proportional to the Ricci scalar of the
induced metric on the horizon (since $\mathbf{\tilde{L}}^{(1)}$ is the
Einstein-Hilbert Lagrangian).  Explicitly, comparing (\ref{gblag}) to
the general form (\ref{lll}) gives
\begin{equation}
\alpha_0 =\frac{(n-1)(n-2)}{16 \pi G n! l^2}, \qquad \alpha_1 =
\frac{1}{16 \pi G (n-2)! }, \qquad \alpha_2 =\frac{\alpha}{16 \pi G
  (n-4)! },
\end{equation}
and so 
\begin{equation}
S = {1 \over 4G} \int_{\Sigma} d^{n-2} x \sqrt{\tilde{h}} \left( 1 +
2\alpha \tilde{R} \right) , 
\end{equation}
where $\tilde{R}$ is the Ricci scalar of the metric
$\tilde{h}_{\mu\nu} = r_+^2 h_{\mu\nu}$ on the bifurcation surface
$\Sigma$.  

Taking the metric $h_{\mu\nu}$ to be a constant-curvature metric, this
will reproduce the formula (\ref{gbent}) for the entropy. We see that
the fact that the correction is proportional to $k$ arises from the
more general statement that the correction to the usual entropy
formula is simply proportional to the integral of the scalar curvature
of the horizon for black hole solutions of this form. This general
connection to horizon curvature was also obtained by the conical
singularity method in~\cite{Furs}.

We can easily extend this result to the case considered
in~\cite{bhscan,Cai:1998,Aros:2000}, where we take the $\alpha_p$ to
be given by (\ref{llalpha}). In this case, we again have that $G(r)$
vanishes on the event horizon, so we can simplify the general formula
for the entropy to
\begin{equation}
S = 2\pi \int_{\Sigma} \sum_{p=1}^{q} p \alpha_p
\mathbf{\tilde{L}}^{(p-1)}.  
\end{equation}
This formula was also obtained by directly integrating the first law
in~\cite{JacMy}. 
If we again take the metric $h_{\mu\nu}$ to have constant curvature,
the curvature of the horizon will be given by
\begin{equation}
\mathbf{\tilde{R}}^{mn}|_{\Sigma} = {k \over r_+^2}
\mathbf{\tilde{e}}^m \mathbf{\tilde{e}}^n, 
\end{equation}
where $k = 0,\pm 1$, and we obtain the entropy
\begin{eqnarray}
S &=& 2\pi \sum_{p=1}^{q} p \alpha_p \left( k \over r_+^2
\right)^{p-1} \int_\Sigma \boldsymbol{\epsilon}_{\tilde h} \\
&=&  \frac{2\pi}{G_q}\sum_{p=1}^{q}
\frac{q!k^{p-1} r_{+}^{n-2p}}{(p-1)!(q-p)!(n-2p)l^{2q-2p}} \\
&=& \frac{2\pi q}{G_q}\sum_{p=0}^{q-1}
\left(\begin{array}{c}q-1\\p\end{array}\right)\frac{k^{p} r_{+}^{n-2p-2}}{(n-2p-2) l^{2q-2p-2}}\\
&=& \frac{2\pi q}{G_q} \int_{0}^{r_{+}} r^{n-2q-1} (k+\frac{
  r^2}{l^2})^{q-1} dr 
\end{eqnarray}
reproducing the expression (\ref{llent}).
Thus, we see that from the Noether charge point of view, the
dependence of the entropy on the the curvature of the horizon, which
appeared somewhat obscure in previous calculations, arises very
naturally.

\section{Negative entropy}
\label{neg}  

We would now like to explore the problem of negative entropy. As
observed in~\cite{Od1,Od2,Cai:2003}, if we
consider the entropy we have assigned to the Gauss-Bonnet black holes
(\ref{gbent}), we see that the entropy is negative if
\begin{equation}
\frac{2 \tilde{\alpha} k
(n-2)}{(n-4) r_{+}^{2}} < -1. 
\end{equation}
This condition can be satisfied for sufficiently small black holes in
either of two ways: if $\tilde \alpha <0$, $k=+1$, or if $\tilde
\alpha >0$, $k=-1$, which latter is possible only for negative
cosmological constant.

Now in both these cases, there is a lower bound on the size of the
black hole---physical black hole solutions do not exist for all values
of $r_+$---so it is still a non-trivial question whether there
actually exist any solutions with negative entropy. 

In the case where $\tilde \alpha <0$, $k = 1$, it was observed
in~\cite{Cai:2001,Cai:2003} that there was a minimum horizon radius,
$r_+^2 > - 2 \tilde \alpha$. For any value of $\alpha$, this still
leaves a finite range of values where the black hole will have
negative entropy:
\begin{equation}
-2\tilde \alpha < r_+^2 < - 2 \tilde \alpha {(n-2) \over (n-4)}.
\end{equation}
In~\cite{Cai:2003}, it was suggested that these solutions be
discarded. However, they have no obvious pathologies as classical
solutions. Therefore, we cannot justify discarding them; the fact that
they appear to have negative entropy should be interpreted as a
problem with our prescription for calculating the entropy. 

In the case where $\tilde \alpha >0$, $k = -1$, which requires
$\Lambda <0$, there is also a minimum size black hole: if we consider
the solution with $M=0$, we can see from (\ref{gbF}) that there is
still a horizon at 
\begin{equation}
r_{+min}^2 = {l^2 \over 2} \left( 1 + \sqrt{1 - {4 \tilde \alpha \over l^2}}
\right).  
\end{equation}
Now this is a vacuum solution, and this horizon is just an
acceleration horizon, corresponding to the fact that the coordinates
we have used do not cover the whole of AdS$_n$. Nonetheless, when we
consider $M>0$, the event horizons of our black hole solutions will
always have $r_+ > r_{+min}$. The question of whether there are any
physical negative entropy solutions is now a little more
non-trivial. The entropy of the solution with horizon radius
$r_{+min}$ vanishes when 
\begin{equation}
\tilde \alpha = l^2 {n (n-4) \over 4(n-2)^2} .
\end{equation}
Thus, for values of $\tilde \alpha$ in 
\begin{equation}
{n (n-4) \over (n-2)^2} {l^2 \over 4} < \tilde \alpha < {l^2 \over 4},
\end{equation}
there are negative-entropy black holes with 
\begin{equation}
r_{+min}^2 < r_+^2 < 2 \tilde \alpha {(n-2) \over (n-4)}.
\end{equation}
The upper bound on the range of $\tilde \alpha$ is the maximum value
for which this theory has a well-defined vacuum
solution~\cite{Cai:2001}.

Since there is a minimum value of the size of the event horizon in the
cases where we are encountering negative entropies, there
is an obvious solution to this problem; we can simply change the lower
limit of integration in (\ref{therment}), defining the entropy to be
\begin{equation}
S = \int_{r_{+min}}^{r_+} T^{-1} \left( {\partial M \over
\partial r_+} \right) dr_+,
\end{equation}
where $r_{+min}$ is the minimum value that the horizon radius takes in
a given family of solutions.\footnote{This additive ambiguity in the
  entropy was also noted in~\cite{Od2}.} This will add an overall
constant to the entropy assigned to all the black hole solutions in a
particular family of solutions. This will reduce to the previous
expression (\ref{therment}) for the usual cases, and will by
construction assign a positive entropy to all the solutions, since the
integrand is positive, so $S$ is an increasing function of $r_+$. This
assigns zero entropy to the solution at $r_+ = r_{+min}$, which has a
non-zero horizon area, which may seem unnatural. However, we saw in
the particular example discussed above that this horizon was just an
acceleration horizon in an AdS$_n$ solution, representing the failure
of our coordinates to cover the whole solution, so this might be a
reasonable assignment. It is certainly more reasonable than ascribing
a negative entropy to some of the solutions.

Thus, the problem of negative entropies can easily be resolved within
this previous framework. However, the main point of our paper was to
re-calculate these entropies from the Noether charge point of view; in
doing so, we have successfully reproduced the sometimes negative
answers of~\cite{Cai:2001,Cai:1998,bhscan}. But in this approach, we
have a purely geometrical expression for the entropy of a particular
solution, not an integral over the family of solutions. So how can we
fix this problem in the Noether charge approach?

In fact, there is a corresponding ambiguity in this
approach\footnote{There is a more obvious ambiguity: the requirement
  that the entropy satisfy the first law would clearly allow us to add
  a universal constant to the definition (\ref{nent}). However, such a term
  can be fixed by requiring that the entropy of flat space
  vanish, and in any case, no single choice of constant could make the entropy
  positive in all families of solutions. The more subtle ambiguity we
  uncover here corresponds to adding a term to (\ref{nent}) which is not a
  universal constant, but depends on the family of solutions.}.  
The Noether charge $\mathbf{Q}$ was defined by $\mathbf{J} =
d\mathbf{Q}$, so it is only defined up to the addition of a closed
form. Previously, the ambiguity has been viewed as one of adding an
exact form $d\mathbf{Z}$, which clearly cannot affect the calculation
of the entropy. However, for our black hole metric (\ref{bhmet}),
there is a natural closed but not exact $(n-2)$-form: the volume form
$\boldsymbol{\epsilon}_h$ associated with the metric $h_{\mu\nu}$.
Note that we must take the $r$-independent metric $h_{\mu\nu}$ rather
than the induced metric $r^2 h_{\mu\nu}$ for this to be a closed form.
The integral of this form over the bifurcation two-surface $\Sigma$ is
clearly non-zero.  Hence, if we redefine $\mathbf{Q}$ by
\begin{equation}
\mathbf{Q}' = \mathbf{Q} - {S_{min} \over \Sigma_k} \boldsymbol{\epsilon}_h, 
\end{equation}
it will change the entropy:
\begin{equation}
S' = S - S_{min}. 
\end{equation}
The coefficient $S_{min}$ must be a constant, independent of the
parameters in a particular solution (but possibly depending on the
parameters $\alpha_p$ in the Lagrangian), so this represents exactly
the same ambiguity---the freedom to shift the entropy of all the black
hole solutions in a particular family of solutions by an overall
constant. The entropy $S'$ will still satisfy the first law, as $S$
does and $\delta S_{min} = 0$.

Why does this ambiguity not appear in discussions of the entropy for
more familiar cases? Here we should note that $\boldsymbol{\epsilon}_h$ is a
well-defined form on the black hole solutions only because the surface
$r=0$, where it might become ill-defined, is not part of the spacetime
manifold. Thus, if we consider a family of solutions which includes a
true vacuum solution, where $r=0$ is a regular point, we lose the
freedom to add such a term. So from this point of view, we could argue
that the Noether charge approach does have one additional piece of
information: it knows we should take $S_{min} =0$ when our family of
solutions has $r_{+min}=0$. But in cases where $r_{+min} \neq 0$,
there is a real ambiguity, and we have to choose $S_{min}$. 

\section{Conclusions}

There are two main points to this paper. First, we illustrated the
application of the Noether charge approach to the black hole
entropy~\cite{Iyer:1994,Iyer:1995} by using it to calculate the
entropy in Gauss-Bonnet and Lovelock theories of gravity. We showed
that if we use the language of differential forms, this provides a
straightforward way to do the calculation, which makes clear the
connection between the corrections to the usual area law and geometric
features of the event horizon. We also observed that there is an
ambiguity in the calculation of the entropy from the first law,
corresponding to a choice of zero of entropy within each family of
black hole solutions. This ambiguity arises in both
the previous approaches to the entropy~\cite{Od2} and in the Noether
charge approach. The additive ambiguity can be used to remove the 
negative entropies which appear in the naive calculations for some of
the families of solutions we considered. This shows that the
appearance of negative entropies is not a sign of some physical
pathology in the corresponding solutions. 

\medskip
\centerline{\bf Acknowledgements}
\medskip    
    
We acknowledge discussions with Christos Charmousis which prompted our
interest in this area. TC was supported in part by a grant from the
Nuffield foundation.  SFR is supported by the EPSRC.

  

\providecommand{\href}[2]{#2}\begingroup\raggedright\endgroup

\end{document}